\documentclass[pdflatex,sn-nature,iicol]{sn-jnl}
\usepackage{graphicx}%
\usepackage{multirow}%
\usepackage{amsmath,amssymb,amsfonts}%
\usepackage{amsthm}%
\usepackage{mathrsfs}%
\usepackage[title]{appendix}%
\usepackage{xcolor}%
\usepackage{textcomp}%
\usepackage{manyfoot}%
\usepackage{booktabs}%
\usepackage{algorithm}%
\usepackage{algorithmicx}%
\usepackage{stfloats}
\usepackage{algpseudocode}%
\usepackage{listings}%
\usepackage{subfigure}
\usepackage{longtable}
\usepackage{tabularx}
\usepackage{threeparttable}

\theoremstyle{thmstyleone}%
\theoremstyle{thmstyletwo}%

\theoremstyle{thmstylethree}%

\begin{document}

\title[Article Title]{Facial Movement Dynamics Reveal Workload During Complex Multitasking}

\author*[1,2]{\fnm{Carter} \sur{Sale}}\email{carter.sale@mq.edu.au}

\author[3]{\fnm{Melissa N.} \sur{Stolar}}

\author[1,2]{\fnm{Gaurav} \sur{Patil}}

\author[1,2]{\fnm{Michael J.} \sur{Gostelow}}

\author[1,2]{\fnm{Julia} \sur{Wallier}}

\author[1,2]{\fnm{Margaret C.} \sur{Macpherson}}

\author[2,4]{\fnm{Jan-Louis} \sur{Kruger}}

\author[2,5]{\fnm{Mark} \sur{Dras}}

\author[3]{\fnm{Simon G.} \sur{Hosking}}

\author[1,2]{\fnm{Rachel W.} \sur{Kallen}}

\author*[1,2]{\fnm{Michael J.} \sur{Richardson}}\email{michael.j.richardson@mq.edu.au}

\affil*[1]{\orgdiv{School of Psychological Sciences}, \orgname{Macquarie University}, \orgaddress{\city{Sydney}, \state{NSW}, \country{Australia}}}

\affil[2]{\orgdiv{Performance and Expertise Research Centre}, \orgname{Macquarie University}, \orgaddress{\city{Sydney}, \state{NSW}, \country{Australia}}}

\affil[3]{\orgdiv{Defence Science and Technology Group}, \orgname{Australian Department of Defence}, \orgaddress{\city{Melbourne}, \state{VIC}, \country{Australia}}}

\affil[4]{\orgdiv{Department of Linguistics}, \orgname{Macquarie University}, \orgaddress{\city{Sydney}, \state{NSW}, \country{Australia}}}

\affil[5]{\orgdiv{School of Computing}, \orgname{Macquarie University}, \orgaddress{\city{Sydney}, \state{NSW}, \country{Australia}}}

\abstract{Real-time cognitive workload monitoring is crucial in safety-critical environments, yet established measures are intrusive, expensive, or lack temporal resolution. We tested whether facial movement dynamics from a standard webcam could provide a low-cost alternative. Seventy-two participants completed a multitasking simulation (OpenMATB) under varied load while facial keypoints were tracked via OpenPose. Linear kinematics (velocity, acceleration, displacement) and recurrence quantification features were extracted. Increasing load altered dynamics across timescales: movement magnitudes rose, temporal organisation fragmented then reorganised into complex patterns, and eye–head coordination weakened. Random forest classifiers trained on pose kinematics outperformed task performance metrics (85\% vs. 55\% accuracy) but generalised poorly across participants (43\% vs. 33\% chance). Participant-specific models reached 50\% accuracy with minimal calibration (2 minutes per condition), improving continuously to 73\% without plateau. Facial movement dynamics sensitively track workload with brief calibration, enabling adaptive interfaces using commodity cameras, though individual differences limit cross-participant generalisation.}


\maketitle

\section{Introduction}\label{sec1}

In safety-critical domains, lapses in performance can have serious consequences \citep{parasuraman_neuroergonomics_2006}. Pilots, air-traffic controllers, and emergency responders must often manage multiple tasks under fluctuating cognitive demands. When these demands outpace an individual's abilities the risk of error rises \citep{masi_stress_2023,zamarreno_suarez_understanding_2024,raduntz_indexing_2020,rubinstein_executive_2001,ophir_cognitive_2009}. As a result, real-time monitoring of operator workload has become a central concern in both research and applied settings \citep{paas_cognitive_2016,korbach_measurement_2017}. 

Despite the growing interest in real-time workload monitoring, widely used assessment methods remain limited in both practicality and resolution. Subjective measures, such as NASA-TLX surveys, and performance-based outcomes are available only after a task is completed or an error has occurred, making them ill-suited for early detection \citep{hart_development_1988,hart_nasa-task_2006}. Additionally, performance outcomes provide limited insight into \textit{how} individuals respond to varying task demands - as prior work has demonstrated, task performance does not consistently align with other indicators of workload \citep{matthews2015psychometrics}.

In response, researchers have turned to physiological signals as more objective indicators of cognitive workload. Measures such as EEG, ECG (e.g., heart rate and heart rate variability), and pupil dilation offer continuous tracking but often rely on specialised hardware and complex setups \citep{das_chakladar_cognitive_2024}. Many biosensors require skin contact or wearable gear, which can interfere with natural behaviour during task execution \citep{mullen_real-time_2015}. Even ``dry'' EEG electrodes and adhesive ECG patches may restrict movement or draw attention due to their physical presence \citep{fu_dry_2020}, and EEG in particular is highly susceptible to electromagnetic interference and extraneous muscle activity—especially outside controlled lab settings \citep{mihajlovic2014wearable}.

Recent efforts have aimed to reduce the intrusiveness of traditional workload monitoring. Remote eye trackers, for example, can capture gaze behaviour without physical contact \citep{hennessey_improving_2009}, and pressure-sensitive textiles have been used to infer posture shifts linked to cognitive strain \citep{qiu_body_2012}. However, these approaches still depend on specialised hardware, limiting their scalability and deployment in naturalistic settings. As a result, there remains a need for low-cost, unobtrusive methods that can operate in everyday environments \citep{heard_survey_2018,das_chakladar_cognitive_2024,almukhtar_objective_2025}. 

An emerging alternative focuses on the person's interaction with the task itself–namely, movement. Research in human factors and ergonomics suggests that cognitive demands can manifest in movement and posture \citep{aitsam_measuring_2025,brunye_movement_2024,mehta2016integrating}. Even coarse measures, such as the amount of motion, have been linked to changes in mental workload \citep{giakoumis_using_2012}, and more recent work has identified consistent patterns of postural adjustment under cognitive strain \citep{adams_work-related_2024,nino2023evaluating}.

Thankfully, with recent advances in computer vision, it is now possible to capture these behavioural signals unobtrusively \citep{stenum2021applications,pagnon2022pose2sim,lahkar2022accuracy}. Markerless pose estimation tools such as OpenPose and DeepLabCut allow researchers to track keypoints on the body using only a standard webcam \citep{cao_realtime_2017,mathis_deeplabcut_2018}. Early results suggest that even simple kinematic features, like velocity or displacement, may reflect cognitive demands \citep{aitsam_measuring_2025}. Unlike eye trackers or biosensors, video-based pose tracking is low-cost, off-body, and does not require calibration, and is easily deployed in naturalistic settings  \citep{albanis_towards_2022,cormier_where_2022,fortini_markerless_2023,choi_human_2024}.

While prior work has largely emphasised movement magnitude or static postural features \citep{giakoumis_using_2012,brunye_movement_2024,aitsam_measuring_2025}, less attention has been given to how the structure of movement evolves over time. Subtle changes in behavioural dynamics may offer a richer window into cognitive strain by capturing not just how much people move, but how that movement is organised.

\begin{figure*}[htbp!]
    \centering
    \includegraphics[width=\linewidth]{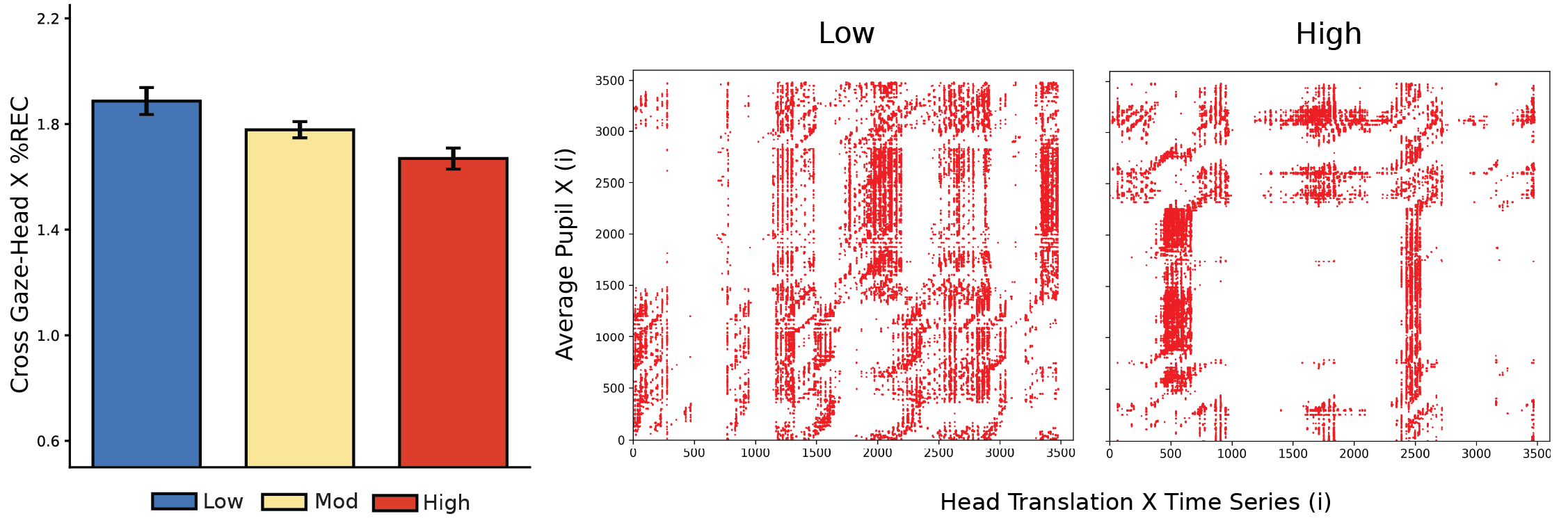}
    \caption{Cross-recurrence between head position and gaze decreases with cognitive load. Left panel: Mean percentage of recurrent points (\%REC) in cross-recurrence quantification analysis (CRQA) between horizontal head movement (head translation X) and horizontal pupil displacement across low, moderate, and high workload conditions. Error bars represent standard error of the mean. Right panels: Representative cross-recurrence plots from a single participant showing the same time window under low (middle) and high (right) cognitive load conditions. Each red point indicates a moment when head and gaze movements exhibited similar dynamical patterns (recurrent states). The denser recurrence structure in the low condition reflects more consistent coordination between head and gaze movements, while the sparser structure in the high condition indicates reduced coupling as task demands increase. Axes show time-delayed embedded dimensions of the respective time series.}
    \label{fig:rec_plots}
\end{figure*}

Recurrence Quantification Analysis (RQA) provides a way to assess this temporal structure. It quantifies how often a system returns to a previous state, producing a recurrence plot that reveals patterns of coordination and stability over time \citep{marwan_recurrence_2007}. For instance, diagonal lines in the plot reflect repeated sequences of behaviour, while vertical lines suggest moments of stagnation \citep{trulla_recurrence_1996} (see Figure~\ref{fig:rec_plots} for examples). 

RQA metrics summarise these patterns: recurrence rate (\%REC) tracks how often states repeat; determinism (DET) captures the proportion of recurrences that form predictable sequences; and entropy measures the complexity of these sequences. Unlike linear metrics that summarise overall magnitude or variance, RQA tracks how structure changes over time - sensitive to subtle disruptions in coordination. This approach has proven useful across domains ranging from cardiac rhythms to interpersonal communication \citep{richardson2007distinguishing,marwan_recurrence_2007,anderson2013recurrence,coco_cross-recurrence_2014,washburn2014dancers,perez2018beyond,brick2018recurrence,crone2021synchronous}. 

What remains unknown is whether these movement dynamics, especially when measured unobtrusively via video-based pose tracking, reflect meaningful variation in cognitive workload during complex multitasking. It also remains an open question whether such pose-derived features can support the classification of workload levels, both within and across individuals. 

\begin{figure*}
    \centering
    \includegraphics[width=\linewidth]{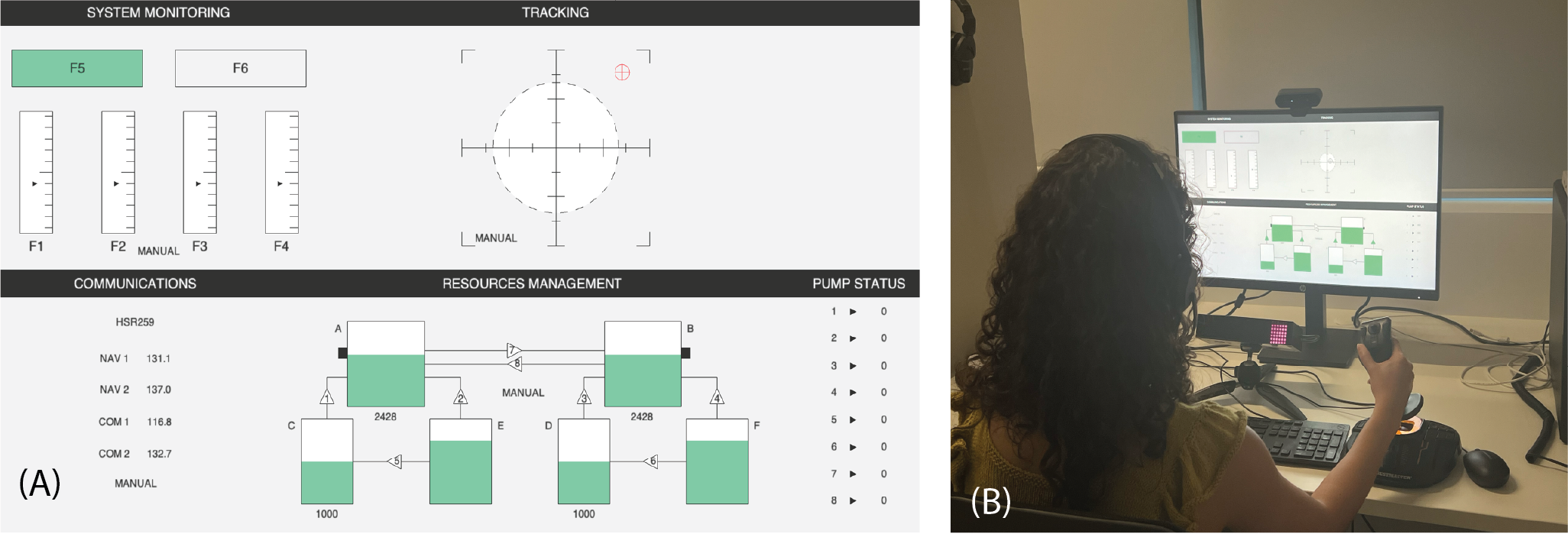} 
    \caption{Overview of the OpenMATB experimental setup. (A) Screenshot of the OpenMATB interface, displaying the four concurrent tasks: System Monitoring, Tracking, Communications, and Resource Management. (B) A participant performing the task using a joystick and monitor setup, with simultaneous recording from a webcam.}
    \label{fig:matb-method}
\end{figure*}

To address this, we applied RQA to head and facial movement data extracted from standard webcam recordings. Participants completed a multitasking protocol based on OpenMATB \citep{cegarra_openmatb_2020}, an open-source implementation of NASA’s Multi-Attribute Task Battery \citep{comstock_jr_multi-attribute_1992}. This environment simulated real-world workload by combining four concurrent subtasks— a continuous tracking task, a system monitoring task, a communications task, and a resource management task (see Figure \ref{fig:matb-method})—under three systematically varied load conditions (Low, Moderate and High), varied across blocks. Each participant completed short (2-minute) baseline blocks at all load levels, followed by longer (8-minute) experimental blocks, allowing for within-subject comparisons and the potential to calibrate models using baseline-derived features.

Pose data were extracted using OpenPose \citep{cao_realtime_2017}, focusing on keypoints associated with attentional and cognitive state - such as head orientation, blink aperture, and gaze movement. From these traces, we computed both linear kinematic features (e.g., velocity, acceleration, RMS displacement) and nonlinear temporal features using RQA (e.g., recurrence rate, determinism, entropy), capturing not only how much participants moved but also how the structure of their movement evolved with task demand.

This design allowed us to assess whether webcam-based facial and upper-body dynamics reflect cognitive workload, and whether they support reliable workload classification. 

By linking moment-to-moment behavioural dynamics to structured variations in task demand, this study evaluates the feasibility of video-based pose tracking as a scalable, non-intrusive method for cognitive state monitoring in complex, multitasking environments.

\section{Results}\label{sec2}

To evaluate whether pose-derived movement features reflected changes in cognitive workload, we examined the data at three levels: task performance, pose-based feature variation, and classification accuracy. Analyses were conducted separately for the short 2-min baseline blocks and the longer 8-min experimental blocks designed to impose sustained demand. First, we examined how task load influenced participants' accuracy and reaction time across subtasks. We then analysed movement features extracted from webcam recordings, including both linear and nonlinear (RQA-based) metrics. Finally, we tested whether these features could support within- and across-participant classification of workload conditions.

\subsection{Cognitive load alters performance differently across subtasks}

\begin{figure*}[ht!]
    \centering
    \includegraphics[width=\linewidth]{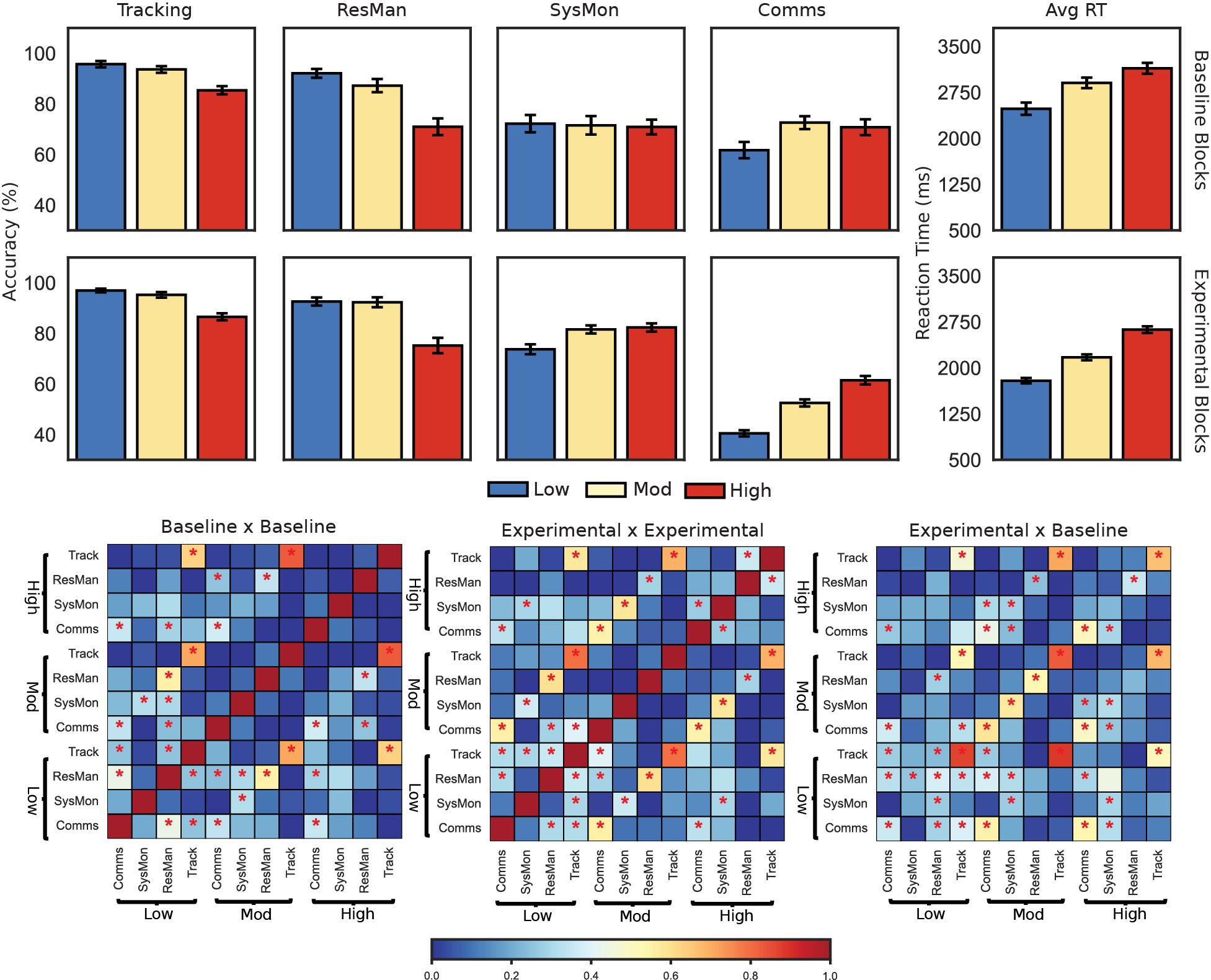} 
    \caption{
        \textbf{Top:} Mean accuracy (\%) and reaction time (ms) across OpenMATB subtasks and load levels for baseline (top row) and experimental (bottom row) blocks. Subtasks include Tracking, Resource Management (ResMan), System Monitoring (SysMon), and Communications (Comms), with average reaction time (Avg RT) shown in the rightmost column. Bars represent means; error bars indicate $\pm1$ SEM. Load conditions are colour-coded: Low (blue), Moderate (yellow), and High (red). \textbf{Bottom:} Pairwise Pearson correlations between subtask performance metrics across load conditions. Red asterisks indicate significant correlations ($p < .05$). Warm colours represent stronger positive correlations.
    }
    \label{fig:performance_and_correlations}
\end{figure*}

To contextualise the movement analyses, we first examined how increasing task demands affected performance across the four subtasks. Rather than a uniform decline, accuracy varied by subtask (see Fig \ref{fig:performance_and_correlations}).  

Tracking and Resource Management both showed reliable accuracy declines with increasing load during the experimental blocks. Tracking accuracy dropped from low to moderate load ($d = -0.3$, $p < .001$) and again from moderate to high ($d = -0.99$, $p < .001$). Resource Management followed a similar trend, with a significant decline from moderate to high load ($d = -0.63$, $p < .001$). Baseline blocks showed comparable patterns, with significant accuracy drops across both load transitions for Resource Management ($p < .05$). 

By contrast, System Monitoring improved slightly under load. Accuracy increased from low to moderate load in the experimental session ($d = 0.47$, $p < .001$), with no change beyond that. Baseline accuracy remained stable across load levels ($p > .911$).

The most consistent gains appeared in the Communications task, where accuracy rose steadily across load levels during the experiment (low-mod $d = 1.14$, $p < .001$; mod-high $d = 0.72$, $p < .001$). This trend appeared in the shorter baseline blocks although the high-load boost did not reach significance (low-mod: $d=0.39,p=0.017$ mod-high: $d = -0.07$, $p = .891$).

Reaction time (RT) was calculated for the two discrete-response tasks and increased with task load. In the experimental blocks, System Monitoring RT rose significantly at both low to moderate ($d=0.37,p<.001$) and from moderate to high load ($d=0.69,p<.001$). Communications followed the same pattern, with increases from low to moderate ($d=1.36,p<.001$) and from moderate to high ($d=1.07,p<.001$). Baseline trends were similar and are reported in the Supplementary Materials 1.1. 

Subjective workload ratings (NASA-TLX), collected after each experimental block, confirmed that the load manipulation was effective. Ratings increased with task load and were negatively associated with performance across subtasks (see Supplementary Materials 1.3).

Together, these findings highlight how increased task demand does not necessarily impair performance. Instead, the effects varied by subtask: while some tasks suffered under load, others improved. This suggests that workload is not simply a function of task difficulty, but reflects how individuals engage with the task.

\subsection{Subtask performance patterns remain stable across time}

To assess whether participants adapted to task demands in stable, individualised ways, we examined whether accuracy in the shorter baseline session predicted accuracy during the longer experimental session.

All subtasks showed significant cross-session correlations, with the strongest effects for tracking ($r = .840$, $p < .001$), overall accuracy ($r = .753$, $p < .001$), and communications ($r = .752$, $p < .001$). Resource management ($r = .489$) and system monitoring ($r = .494$) were also significant (both $p < .001$). Removing outliers (3 SD) reduced most correlations but preserved the general pattern, except that system monitoring was no longer significant (see Supplementary Materials 1.4). Overall, individual differences in task performance were largely stable across sessions, with greater consistency in continuous tasks than in discrete ones.

To visualise these patterns, we computed pairwise correlations between accuracy scores across all load levels and sessions (Figure~\ref{fig:performance_and_correlations}). Within-session matrices (left and centre panels) revealed the strongest off-diagonal coupling between the low and moderate load conditions. This suggests that subtask performance patterns were more alike between these two levels, implying a similar degree of task demand. 

In the Experimental session (centre panel), cross-subtask consistency extended more visibly into the High-load condition compared to Baseline, indicating that with practice, participants were better able to integrate and manage multiple tasks even under elevated demands. 

The cross‐session matrix (right panel) shows a statistically-robust diagonal for each task and load, confirming that subtask-performance patterns remained stable across time (between experimental and baseline blocks).

These findings build on the earlier observation that rising workload does not uniformly degrade performance. Instead, different subtasks respond differently to increasing cognitive demands–some showing declines and others improving. These patterns held across both baseline and experimental blocks, suggesting that performance is shaped not just by how much load is applied, but by the specific demands of each subtask. This consistency across sessions suggests that individuals may engage with tasks in stable, characteristic ways.

\subsection{Facial and gaze kinematics vary with task load}

Having established that increased task demands influence performance in subtask-specific ways, we next examined whether those differences are reflected in the way the person interacts with the task–movement. 

Four feature regions were extracted from the pose data: (1) head movement (translation, rotation, and scale parameters from Procrustes alignment), (2) pupil motion (estimated from eye landmark positions), (3) blink aperture (Euclidean distance between upper and lower eyelids), and (4) mouth opening (Euclidean distance between upper and lower lips). For each feature, displacement, velocity, and acceleration time series were computed. The time series were then divided into 60-second windows (50\% overlap) and summary statistics were extracted. We focus primarily on root mean square (RMS) and mean values in the main text, though all nine statistics (including standard deviation, median, minimum, maximum, percentiles, and autocorrelation) were used in machine learning models. For translational features (head position, pupils), we report both per-axis components ($X,Y$) and the 2D magnitude (Euclidean); rotation, blink, and mouth aperture are scalars. The full feature extraction procedure is described in the Methods section. We focus on experimental blocks here; baseline results followed similar trends and complete statistical results for all features can be found in Supplementary Materials 2.1-2.2.

Head movement showed increased velocity and acceleration primarily at high load. While overall head motion magnitude showed minimal displacement changes, velocity increased substantially from moderate to high load ($d=0.35,p<.001$), with acceleration following a similar pattern ($d=0.25,p<.001$). This suggests that while the extent of head movement remained relatively stable, participants moved their heads more quickly when demands peaked.

Mean head rotation increased steadily across both transitions (L--M: $d=0.13,p<.001$; M--H: $d=0.15,p<.001$), indicating a progressive shift in head orientation. However, rotation velocity increased only at high load (M-H: $d=0.38,p<.001$), with acceleration showing a similar effect (M-H: $d=0.26,p<.001$). This suggests that under moderate load, participants repositioned to new viewing angles but did so smoothly, whereas high load demanded rapid switching between orientations to keep pace with multiple concurrent events. 

Vertical head position shifted progressively downward, particularly at high load (M--H: $d=-0.12,p<.001$), with velocity (M--H: $d=0.36,p<.001$) and acceleration (M--H $d=0.23,p<.001$) both increasing. Horizontal translation showed similar velocity and acceleration effects at high load (velocity M-H: $d=0.40,p<.001$; acceleration M-H $d=0.31,p<.001$). This downward drift likely reflects participants increasingly looking down towards the keyboard to respond to task demands. 

Pupil movements showed the strongest and most consistent load effects, particularly for vertical motion. Horizontal pupil RMS decreased slightly from low to moderate load ($d = -0.11$, $p < .001$), while mean position shifted significantly at high load (M--H: $d = 0.32$, $p < .001$). Vertical pupil mean position showed large effects at high load (M--H: $d = 0.30$, $p < .001$). Vertical velocity RMS showed minimal change from low to moderate load ($d = 0.01$, $p = .966$) but increased substantially at high load (M--H: $d = 0.45$, $p < .001$), with acceleration RMS following the same pattern (M--H: $d = 0.42$, $p < .001$). Horizontal velocity RMS and acceleration RMS also increased primarily at high load (velocity M--H: $d = 0.37$, $p < .001$; acceleration M--H: $d = 0.37$, $p < .001$). Overall pupil magnitude velocity RMS increased primarily from moderate to high load (M--H: $d = 0.39$, $p < .001$), with acceleration RMS showing similar effects (M--H: $d = 0.37$, $p < .001$). Pupil metric RMS (overall pupil diameter variability) decreased with load (L--M: $d = -0.09$, $p = .008$), as did mean pupil metric (L--H: $d = -0.17$, $p < .001$). 
These patterns likely reflect both increased vertical scanning as participants looked down toward task-relevant display regions and more rapid shifts in gaze under peak task demand.

\begin{figure}[htp!]
    \centering
    \includegraphics[width=\linewidth]{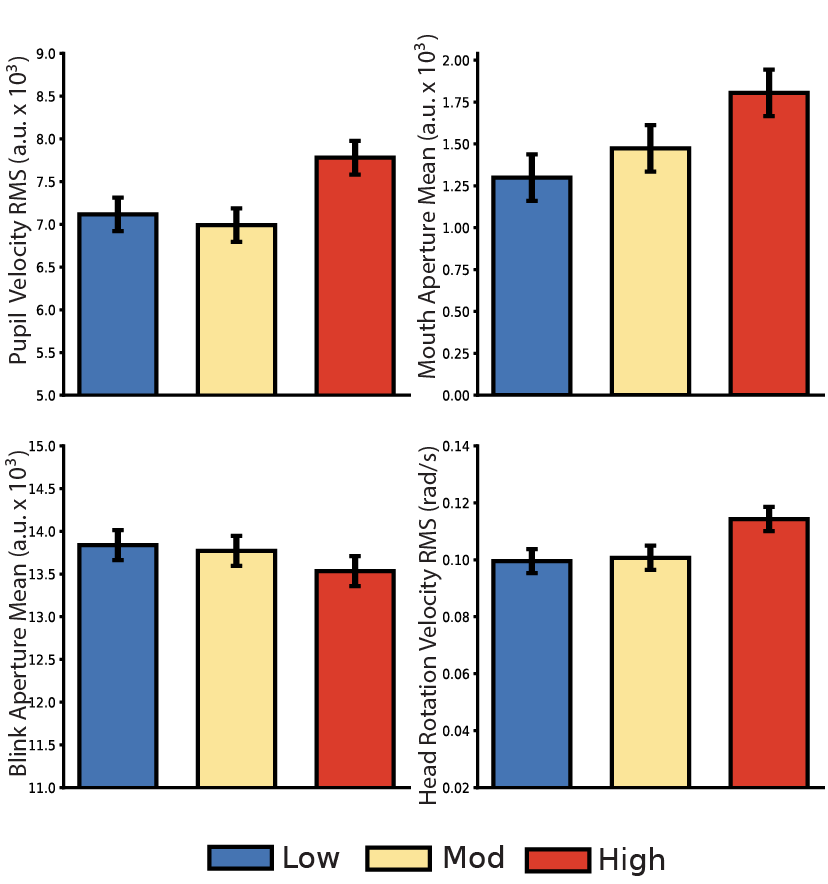}
    \caption{Bar plots show mean values across load conditions (Low, Moderate, High) for four key facial movement features. \textbf{Top-left:} Pupil Velocity (RMS) increased progressively, particularly from moderate to high load, reflecting increased vertical visual scanning. \textbf{Top right:} Mouth aperture (mean) increased steadily across both load transitions, consistent with greater communication task demands. \textbf{Bottom left:} Blink aperture (mean) decreased progressively, indicating narrowed eye opening under load. \textbf{Bottom right:} Head rotation velocity (mean) showed a threshold effect, increasing only at high load. Error bars represent $\pm1$ SEM.} 
    \label{fig:linear_bars}
\end{figure}

Blink aperture decreased progressively while blink dynamics intensified. Blink aperture RMS declined from low to high load ($d = -0.22$, $p < .001$), with mean aperture showing an even stronger decline from moderate to high load 
(M--H: $d = -0.26$, $p < .001$), indicating that participants' eyes remained more narrowed as task demands increased. Simultaneously, blink velocity RMS increased progressively at both transitions (L--M: $d = 0.21$, $p < .001$; M--H: $d = 0.39$, $p < .001$), with acceleration RMS following the same pattern (L--M: $d = 0.18$, $p < .001$; M--H: $d = 0.34$, $p < .001$). This pattern likely reflects an increased blink rate under load.

Mouth aperture increased consistently across all kinematic measures. Both RMS and mean aperture grew progressively at both load transitions (RMS L--M: $d = 0.21$, $p < .001$; M--H: $d = 0.33$, $p < .001$; mean L--M: $d = 0.19$, $p < .001$; M--H: $d = 0.37$, $p < .001$). Velocity RMS showed large, consistent increases (L--M: $d = 0.34$, $p < .001$; M--H: $d = 0.33$, $p < .001$), as did acceleration RMS (L--M: $d = 0.33$, $p < .001$; M--H: $d = 0.34$, $p < .001$). These likely reflect increased response rates to the communications task.

Overall, facial and gaze-related kinematics increased with cognitive load, but the timing and nature of these changes varied across features (Fig~\ref{fig:linear_bars}). Pupil movements, blinks, and mouth activity scaled progressively with demand. In contrast, head velocity and acceleration effects, emerged only when task pace outstripped smooth control. Importantly, these movement patterns align with the task structure itself: vertical gaze shifts matched the display layout, mouth activity tracked communication demands, and head adjustments reflected the need to monitor spatially distributed subtasks. These task-aligned changes provide a promising feature space for workload inference, forming the basis for the classification analyses described next. Before turning to that, we examine whether the temporal organisation of these signals–beyond movement magnitude–was also shaped by task demands.

\subsection{Facial dynamics show fragmentation followed by reorganisation}

While linear metrics revealed that movement increased with load, they did not address how those movements are organised. Recurrence quantification analysis (RQA) probes the temporal dynamics of signals, capturing qualities like repetitiveness and predictability. Here, we use RQA to examine the evolution of head motion, pupil motion, head rotation, blink, and mouth dynamics across load conditions.

Rising task demands produced a consistent two-phase pattern across most features: initial fragmentation under moderate load, followed by reorganisation at high load. The primary exception was mouth dynamics, which showed progressive structuring throughout. 

\paragraph{Fragmentation under moderate load}
From low to moderate load, most features showed declining recurrence and determinism, indicating that movement patterns became less repetitive and predictable. Pupil magnitude recurrence decreased ($d = -0.12$, $p = .021$) alongside substantial drops in determinism ($d = -0.28$, $p < .001$) and laminarity ($d = -0.31$, $p < .001$). Both pupil dimensions fragmented similarly: vertical pupil motion showed declining recurrence ($d = -0.12$, $p = .017$) and determinism ($d = -0.11$, $p = .024$), while horizontal motion exhibited comparable fragmentation in determinism ($d = -0.13$, $p = .006$) and laminarity ($d = -0.15$, $p = .002$). Head motion magnitude exhibited declining recurrence ($d = -0.13$, $p = .007$), and head rotation showed similar patterns ($d = -0.15$, $p = .001$). Head translation along both axes also fragmented (T\textsubscript{x}: $d = -0.16$, $p < .001$; T\textsubscript{y}: $d = -0.18$, $p < .001$). Blink dynamics remained relatively stable during this initial transition, with minimal changes in recurrence ($d = -0.04$, $p = .635$) or determinism ($d = 0.02$, $p = .887$).

\paragraph{Reorganisation at high load}
From moderate to high load, determinism increased dramatically across most features, accompanied by substantial increases in entropy---indicating that while systems reorganised into more structured patterns, these patterns were also more complex. Pupil magnitude showed the clearest reorganisation: determinism increased substantially ($d = 0.38$, $p < .001$) despite recurrence remaining suppressed, and entropy rose sharply ($d = 0.34$, $p < .001$). 
Laminarity---reflecting sustained periods in specific states---increased ($d = 0.40$, $p < .001$), as did mean line length ($d = 0.30$, $p < .001$), indicating longer sequences of repeated behaviour.  

The two pupil dimensions diverged at high load. Vertical pupil motion exhibited pronounced reorganisation: determinism increased substantially ($d = 0.33$, $p < .001$), alongside increases in entropy ($d = 0.34$, $p < .001$), laminarity ($d = 0.35$, $p < .001$), and mean line length ($d = 0.29$, $p < .001$). This suggests that vertical gaze patterns, initially disrupted by moderate load, adopted new, more structured sequences under peak demand---likely reflecting systematic downward scanning aligned with the vertical task layout. In contrast, horizontal pupil motion showed no reorganisation and remained fragmented, with determinism showing no recovery ($d = -0.09$, $p = .116$) and laminarity declining further ($d = -0.11$, $p = .034$). This dimension-specific response suggests participants narrowed their horizontal exploration and adopted structured vertical search strategies.

\begin{figure}
    \centering
    \includegraphics[width=\linewidth]{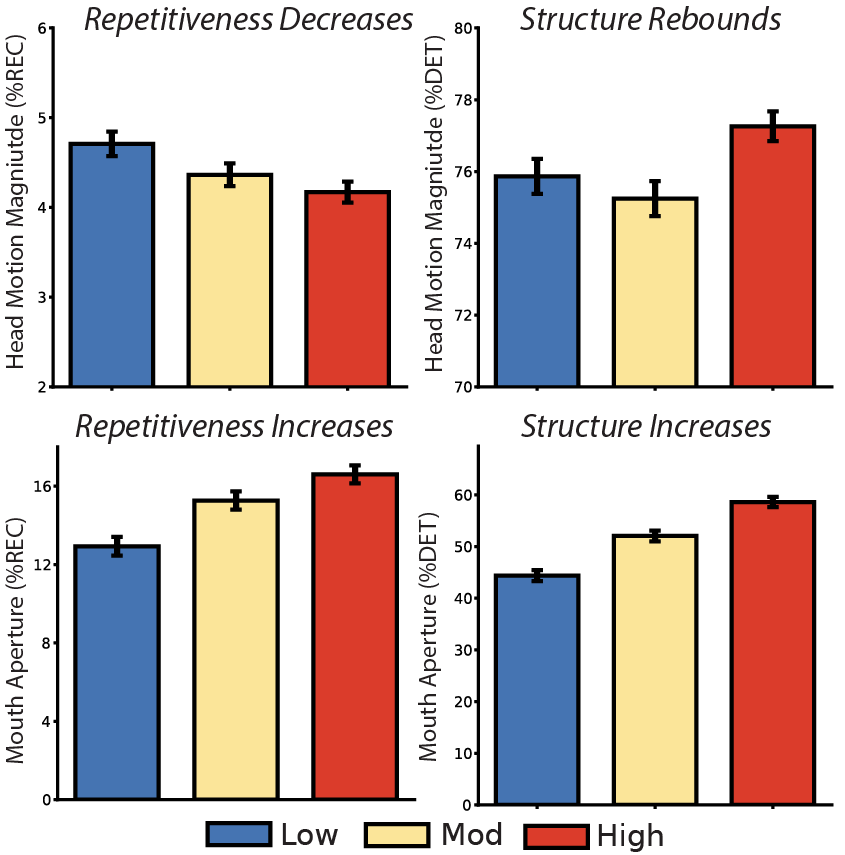}
    \caption{Head and mouth dynamics show opposite trajectories in recurrence (\%REC) and determinism (\%DET) across load levels. From low to moderate load, head motion became less repetitive and predictable, reflecting fragmentation of movement patterns. At high load, head motion determinism rebounds, indicating reorganisation into more structured dynamics. In contrast, mouth aperture becomes increasingly repetitive and structured with rising load. Bars represent means $\pm$ SEM.}
    \label{fig:placeholder}
\end{figure}

Head movements showed similar reorganisation patterns. Head motion magnitude determinism increased at high load ($d = 0.20$, $p < .001$), with corresponding increases in entropy ($d = 0.16$, $p < .001$) and laminarity ($d = 0.19$, $p < .001$). Head rotation determinism rose substantially ($d = 0.19$, $p < .001$), as did entropy ($d = 0.19$, $p < .001$) and laminarity ($d = 0.15$, $p = .001$). Head translation showed particularly large reorganisation effects, with horizontal translation (Tx) determinism increasing ($d = 0.33$, $p < .001$) alongside entropy ($d = 0.35$, $p < .001$) and laminarity ($d = 0.28$, $p < .001$). Vertical translation (Ty) showed comparable increases (determinism: $d = 0.21$, $p < .001$; entropy: $d = 0.15$, $p = .002$; laminarity: $d = 0.20$, $p < .001$).

Blink dynamics diverged from the typical two-phase pattern, showing minimal change under moderate load but substantial structuring at high load. Determinism increased ($d = 0.29$, $p < .001$), accompanied by large increases in entropy ($d = 0.34$, $p < .001$), laminarity ($d = 0.26$, $p < .001$), and mean line length ($d = 0.32$, $p < .001$). Recurrence also increased modestly ($d = 0.15$, $p = .002$). This suggests that blink patterns remained relatively unaffected until demands reached their peak, at which point they reorganised into more structured, periodic sequences.

\paragraph{Progressive structuring of mouth dynamics}

In contrast to other features, mouth aperture showed consistent, progressive increases in structure across both load transitions, with no initial fragmentation phase. Recurrence increased steadily (L--M: $d = 0.16$, $p < .001$; M--H: $d = 0.11$, $p = .024$), as did determinism (L--M: $d = 0.27$, $p < .001$; M--H: $d = 0.23$, $p < .001$), entropy (L--M: $d = 0.20$, $p < .001$; M--H: $d = 0.19$, $p < .001$), and laminarity (L--M: $d = 0.28$, $p < .001$; M--H: $d = 0.23$, $p < .001$). Maximum line length also grew substantially at high load ($d = 0.17$, $p < .001$). This progressive structuring likely reflects the increasing frequency and regularity of verbal responses demanded by the communications task, which scaled directly with load level.

These results reveal a two-phase adaptive response to task demands. Under moderate load, most facial movement systems initially fragment—becoming less repetitive and predictable as individuals adjust to increased task demands. However, at high load, rather than continued degradation, systems reorganise into new coordinative patterns characterised by increased determinism, entropy, and complexity. This suggests that peak cognitive demands induce qualitatively different dynamical regimes: movements become simultaneously more structured (higher determinism, longer sequences) and more complex (higher entropy), possibly reflecting the emergence of compensatory strategies or stereotyped responses. Mouth dynamics, however, scaled progressively with communication demands, highlighting task-specific structuring that aligned directly with subtask requirements. While individual subsystems showed this adaptive reorganisation, the coordination between these systems tells a different story. 

\subsection{Load disrupts gaze-head coupling}

In natural settings, gaze and head movements are typically tightly coupled, reflecting integrated control of visual attention. To test whether this coordination degrades under strain, we examined cross-recurrence between gaze and head movement signals across load conditions.

Coupling between gaze direction and head movement weakened progressively as task load increased. For the magnitude measure (combining both axes), percentage recurrence declined across both transitions (L--M: $d = -0.14$, $p = .005$; M--H: $d = 0.02$, $p = .841$), indicating fewer returns to previously shared gaze--head states. Maximum line length showed consistent reductions (L--M: $d = -0.17$, $p < .001$; M--H: $d = 0.06$, $p = .369$), while mean line length also declined from low to moderate load ($d = -0.15$, $p = .002$) before stabilising. Other measures---including standard line length and maximum vertical line length---showed similar patterns, consistent with reduced coordination stability.

Unlike the individual facial features, cross-recurrence showed no compensatory reorganisation. Determinism declined initially from low to moderate load ($d = -0.13$, $p = .009$) with no recovery ($d = 0.05$, $p = .450$), contrasting sharply with the dramatic increases seen in individual systems. Entropy decreased from low to moderate load ($d = -0.21$, $p < .001$) and showed only modest recovery at high load ($d = 0.11$, $p = .026$), far less robust than the entropy increases that characterised individual feature reorganisation. Trapping time---reflecting sustained periods of coordination---also declined (L--M: $d = -0.12$, $p = .021$; M--H: $d = -0.08$, $p = .185$).

Axis-specific patterns revealed additional detail. Along the X-axis, recurrence declined significantly across both transitions (L--M: $d = -0.09$, $p = .110$; M--H: $d = -0.08$, $p = .136$), and maximum line length decreased (L--M: $d = -0.10$, $p = .060$; M--H: $d = 0.00$, $p = .997$). The Y-axis showed a more pronounced pattern: recurrence declined across both transitions (L--M: $d = -0.13$, $p = .008$; M--H: $d = -0.04$, $p = .618$), accompanied by reductions in maximum line length (L--M: $d = -0.18$, $p < .001$; M--H: $d = 0.06$, $p = .311$) and mean line length (L--M: $d = -0.17$, $p < .001$). However, determinism showed a different trajectory---decreasing initially ($d = -0.16$, $p < .001$) then rebounding substantially at high load ($d = 0.16$, $p < .001$), while entropy followed the same U-shape (L--M: $d = -0.15$, $p = .001$; M--H: $d = 0.14$, $p = .004$). Laminarity also showed this pattern (L--M: $d = -0.14$, $p = .006$; M--H: $d = 0.13$, $p = .006$).

The results indicate distinct patterns of reorganisation within individual facial systems and their coordination. Most individual systems exhibited systematic adaptation under load, characterised by initial fragmentation followed by increased structural integration and complexity. In contrast, the coordination between gaze and head movements showed a progressive breakdown in overall coupling strength, with only the vertical axis showing evidence of late-stage reorganisation in temporal structure. Thus, while individual subsystems reorganise under constraint, inter-system coordination does not necessarily exhibit a comparable response.

\subsection{Pose features outperform task metrics in classifying workload}

\begin{figure*}[htbp!]
    \centering
    \includegraphics[width=\linewidth]{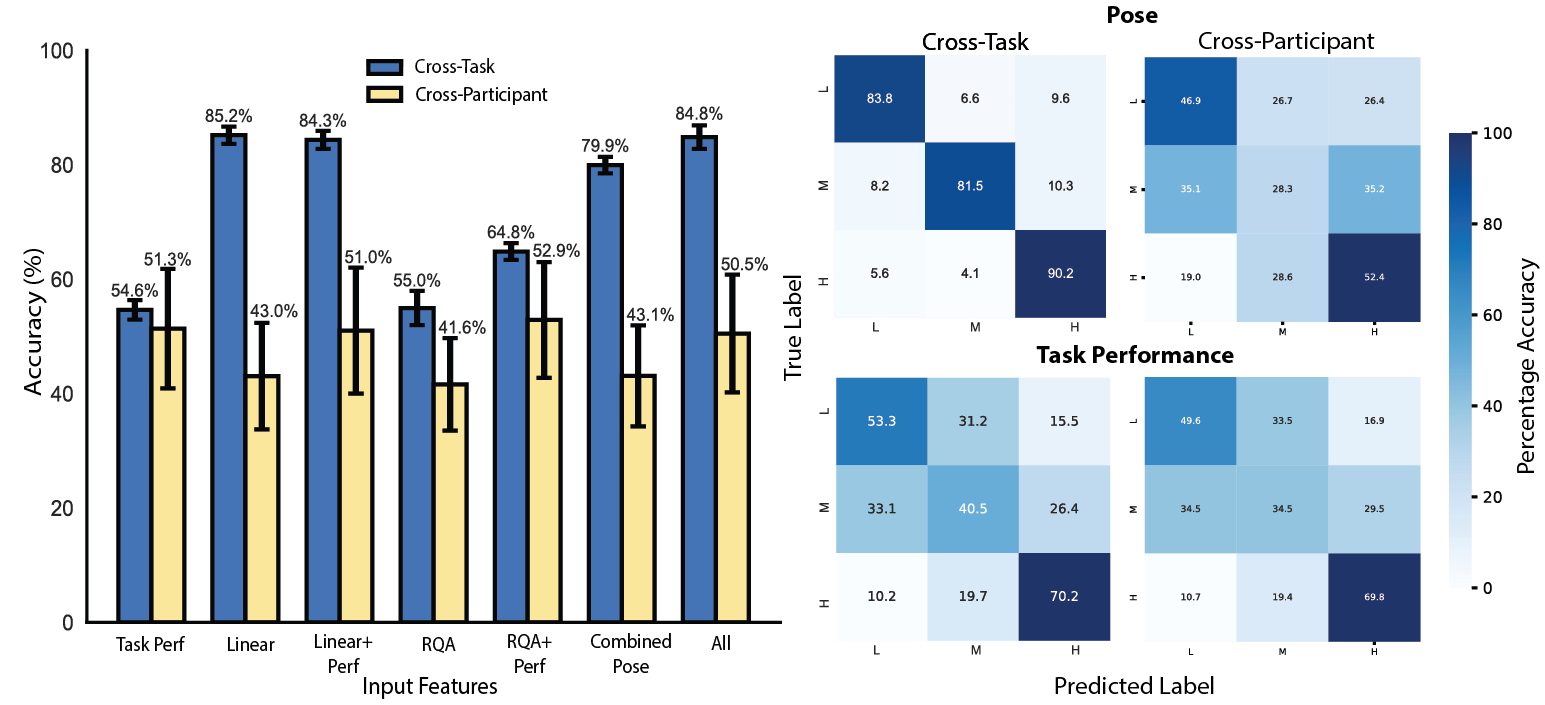}
    \caption{Balanced accuracy (left) for models trained with different feature sets (Linear, Nonlinear, Combined, Task Performance, and their combinations) under Cross-Task and Cross-Participant validation. Percentages above bars indicate mean accuracy across 15 random seeds; error bars show standard deviation. Confusion matrices (right) for the Random Forest model trained on linear kinematic pose features (top) and task performance features (bottom) under Cross-Task (left) and Cross-Participant (right) validation. Cell values indicate class-wise prediction percentages, averaged over 15 seeds.}
    \label{fig:model_perf_bars}
\end{figure*}

The preceding analyses showed that facial and gaze features vary systematically with task load, both in magnitude and temporal organisation. We next asked whether these features were informative enough to classify load condition.

To test this, we trained machine learning models to classify load level (Low, Moderate, High) using randomly sampled task windows (60 seconds with 50\% overlap; “cross-task”). Models used backward feature selection averaged across 15 random seeds (see Methods for details). We report accuracy as the primary performance metric; additional results ($F_1$ and $\kappa$) are provided in Supplementary Material 5.1.

A model trained on all pose features (kinematics and recurrence) achieved 79.9\% $\pm$ 1.5\% accuracy. In contrast, a model using only task performance (accuracy and reaction time) reached just 54.6\% $\pm$ 1.7\% accuracy (Figure~\ref{fig:model_perf_bars}).

To assess which features were driving performance, we trained models using kinematic and recurrence features separately. Linear kinematics alone achieved 85.2\% $\pm$ 1.5\%, outperforming the combined model, while recurrence features achieved only 55.0\% $\pm$ 3.0\%. 

To test whether task performance metrics complemented pose-derived features, we trained additional models combining both feature sets (see Figure~\ref{fig:model_perf_bars}). In the cross-task validation, adding performance features provided no benefit. Linear kinematics combined with task performance achieved 84.3\% $\pm$ 1.6\%, within the confidence interval of linear features alone (85.2\% $\pm$ 1.5\%). Similarly, models combining all pose features with task performance (84.8\% $\pm$ 2.0\%) showed no improvement over linear-only models. The only combination showing modest gains was recurrence plus task performance (64.8\% $\pm$ 1.5\%), which remained substantially below linear kinematics alone.

These results suggest that linear pose metrics capture sufficient workload-relevant structure for within-participant classification, with recurrence features and task performance measures contributing little additional predictive power in this context. Recurrence features remain valuable for revealing \textit{how} workload shapes the temporal structure and dynamics of behaviour, even if they contribute little to raw predictive accuracy.

\subsection{Individual differences limit generalisation}

We next tested how well these features generalised to new individuals using leave-one-participant-out cross-validation, training on n-1 participants and testing on the held-out participant, repeated for all 72 participants (see Methods).

Task performance features achieved 51.3\% $\pm$ 10.4\% accuracy. A model trained on all pose features (kinematics and recurrence) reached only 43.1\% $\pm$ 8.8\% accuracy. When tested separately, recurrence slightly underperformed linear kinematics recurrence features (41.6\% $\pm$ 8.1\% vs.\ 43.0\% $\pm$ 9.3\%), but both showed poor generalisation (Figure~\ref{fig:model_perf_bars}).

Unlike the cross-task scenario, adding task performance to pose features in cross-participant models produced modest improvements. Linear kinematics plus task performance achieved 51.0\% $\pm$ 11.0\%, exceeding linear features alone (43.0\% $\pm$ 9.3\%). Similarly, recurrence plus task performance reached 52.9\% $\pm$ 10.1\%, the highest cross-participant accuracy, compared to 41.6\% $\pm$ 8.1\% for recurrence alone. However, the all-features model (50.5\% $\pm$ 10.3\%) performed comparably to task performance alone (51.3\% $\pm$ 10.4\%), suggesting diminishing returns when combining all feature types.

These results highlight the challenge of cross-participant generalisation: while pose features clearly encode load within individuals, they do not transfer reliably across them. Task performance generalised better, though remained well below ceiling—suggesting that participants engage with the task in idiosyncratic ways. The modest benefit of combining pose and performance features in cross-participant settings suggests these signals capture partially independent aspects of workload expression that vary across individuals.

\subsection{Load prediction emerges with minimal task data}

\begin{figure*}
    \centering
    \includegraphics[width=\linewidth]{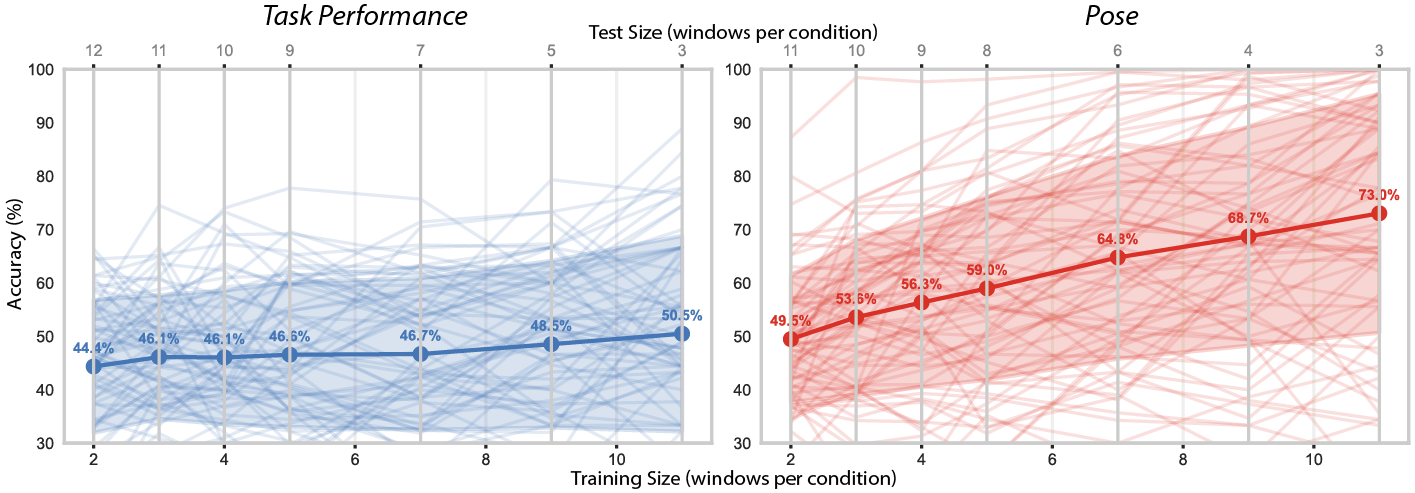}
    \caption{Participant-specific models were trained with increasing amounts of experimental data (2-11 60-second windows per condition, bottom x-axis) and evaluated on held-out test windows (top x-axis, grey). Thin lines show individual participants; thick lines show population means with shaded $\pm1$ SD. (Left) Task performance metrics (accuracy and reaction time across MATB subtasks) showed minimal improvement with additional training data, plateauing around 45-51\% balanced accuracy. (Right) Pose kinematics (facial and head movement velocity, acceleration, displacement) improved continuously throughout the tested range, reaching 73\% accuracy at 11 windows per condition with no evidence of plateau. Individual variability was considerable (SD $>$ 20\% at later training sizes), with some participants achieving $>$80\% accuracy early while others required more extensive calibration. Numbers above points indicate population mean accuracy.}
    \label{fig:learning_curves}
\end{figure*}

Finally, we tested how much task data was required to achieve reliable classification. We trained participant-specific models using varying amounts of training data (2-11 windows per condition, with each window representing 60 seconds of data). Models were evaluated across 10 random seeds per participant to ensure robust estimates. Full per-minute results are reported in Appendix 4.2.

Pose kinematics (facial and head movement velocity, acceleration, displacement) achieved markedly higher accuracy than task performance metrics (subtask accuracy, reaction times) across all training sizes (Figure ~\ref{fig:learning_curves}). At minimal calibration (2 windows per condition), pose features achieved 49.5\% $\pm$ 11.8\% accuracy compare to 44.4\% $\pm$ 12.4\% for performance metrics. This gap widened substantially with additional training data: at 11 windows, pose features reached 73.0\% $\pm$ 22.4\% accuracy versus 50.5\% $\pm$ 18.3\% for performance metrics–a 22.5 percentage point advantage. 

Combining pose and performance features provided no benefit over pose features alone (49.1\% vs 49.5\% at 2 windows; 74.3\% vs 73.0\% at 11 windows), and in some cases slightly degraded performance. This suggests that the performance data introduced variance misaligned with the evolving coordination structure. Similarly, adding recurrence quantification analysis (RQA) features provided no benefit and slightly reduced accuracy at larger training sizes (72.5\% vs 73.0\% at 11 windows). Together, these results indicate that simple linear kinematic features capture the workload-relevant structure in movement dynamics without requiring more complex temporal or task-level measures.

We tested whether including a 2-minute pre-task baseline improved predictions on experimental data. Models trained exclusively on baseline data (training size 0) performed near chance when tested on experimental sessions (37.2\% $\pm$ 10.6\% for pose features; 48.0\% $\pm$ 11.8\% for performance metrics), highlighting the the contextual sensitivity of load expression: even subtle shifts in task structure or engagement substantially reorganise movement behaviour.

When baseline was combined with experimental data, the benefit was negligible. For pose features, adding baseline calibration reduced accuracy by 2.8 percentage points at 2 windows (46.7\% vs. 49.5\%) and 2.7 points at 11 windows (70.3\% vs. 73.0\%). This pattern held across all training sizes, indicating that within-task calibration data alone is sufficient for participant-specific workload detection. 

Learning curves showed steady improvement across all training sizes tested (Figure ~\ref{fig:learning_curves}). For pose features, performance increased continuously from 49.5\% $\pm$ 11.8\% at 2 windows per condition to 73.0\% $\pm$ 22.4\% at 11 windows per condition, with gains of 8.2 percentage points between 7 and 11 windows. The learning curve showed no evidence of plateau at the maximum training size tested, suggesting that additional calibration data would yield further improvements. Performance metrics, by contrast, showed minimal learning across training sizes, improving only 6.1 percentage points from 2 to 11 windows and plateauing around 50\% accuracy.

Individual differences in calibration requirements were considerable. At 11 windows per condition, accuracy ranged from 23\% to 98\% across participants (Figure ~\ref{fig:learning_curves}), with standard deviations exceeding 20 percentage points at later training sizes. This variability underscores both the necessity of participant-specific calibration and the heterogeneity in how individuals express workload through movement dynamics.

Together, these findings show that rich load-related structure emerges quickly within the task itself, even with minimal training data. While baseline behaviour does not reliably transfer, and performance metrics offer limited added value, real-time pose features provide a dense and rapidly accessible signal. Generalisation across contexts remains a challenge—but within-task classification can reach moderate accuracy (approximately double chance performance) with just a few minutes of per-user data, and continues to improve with additional calibration.

\section{Discussion}\label{sec12}

Our results show that movement dynamics change at multiple levels in response to task demands, from shifts in basic kinematic measures to alterations in deeper temporal structure. These changes were consistent across conditions and could be captured with a simple low-cost webcam, demonstrating that complex workload-related dynamics can be measured without specialised or invasive equipment \citep{fortini_markerless_2023}.

Changes in temporal structure generally supported earlier reports of load-related fragmentation in gaze behaviour \citep{kim_applying_2017,yang_measuring_2019}, but here that fragmentation was followed by reorganisation at higher load. Most signals (face, eyes, head, blinks) became less repetitive yet more organised and complex, while mouth movements grew steadily more structured as load increased—evidence of reorganisation rather than simple breakdown.

How do these movement changes relate to overt task outcomes? In our data, higher load did not uniformly impair performance across subtasks. Continuous control tasks (tracking, resource management) tended to decline with load, whereas discrete decision tasks (system monitoring, communications) showed small improvements; reaction times nevertheless lengthened in the discrete tasks as demand increased. Cross-session correlations further suggested that participants expressed these performance patterns in stable, individualised ways. Together, this variability cautions against treating performance as a direct proxy of load and motivates a focus on movement, which more immediately reflects the organisation of task engagement \citep{devlin2020transitions,howard2021using,aitsam_measuring_2025,richer_machine_2024,andersson_effect_2002}.

Beyond performance, workload clearly altered movement kinematics: facial expressions, eye movements, and blinks became faster and more forceful (higher velocities and accelerations) as demand rose. This diverges from reports of reduced gross body motion under load \citep{kannape_cognitive_2014,malcolm_cognitive_2018,farvardin_effects_2022,hagio_effects_2020,aitsam_measuring_2025}, likely reflecting differences in behavioural scale - our fine-grained features (gaze shifts, head adjustments, facial actions) capture rapid micro-adjustments that can increase under strain.

Recurrence analyses further showed that, with higher demand, movement sequences were shorter-lived and less repetitive, and coordination between normally coupled systems weakened: eye–head coupling declined with load, consistent with evidence that added cognitive demand disrupts integrated motor coordination \citep{lustig_higher_2023}. Most facial systems exhibited initial fragmentation followed by reorganisation—reduced recurrence at moderate load, then increased determinism and entropy at high load. This trajectory of increasing complexity contrasts with reports that load drives behaviour toward simpler, more stereotyped patterns \citep{herrebroden_cognitive_2023}. Eye–head coordination, however, showed progressive breakdown without recovery, while mouth dynamics became steadily more structured. Overall, rising task demands produced movement patterns that were more complex and reorganised, rather than a case of simple degradation.

From a modelling standpoint, classifiers trained on pose features captured robust, within-person signatures of workload but transferred poorly across people—consistent with idiosyncratic engagement patterns \citep{lobo_cognitive_2016,heard_survey_2018,xiong_pattern_2020}. Linear kinematic features alone achieved 85\% balanced accuracy in within-participant validation, substantially outperforming task performance metrics (55\%) and recurrence features (55\%). Notably, combining pose features with task performance provided no benefit: models using both feature types achieved 84\%–85\% accuracy, indistinguishable from pose features alone. This finding is particularly striking given that task performance is often considered a primary workload indicator \citep{wickens_multiple_2008,young_state_2015}. The results suggest that pose dynamics capture workload-relevant structure that is either more informative than, or orthogonal to, traditional performance measures—and that performance metrics may introduce noise when combined with kinematic features.

Cross-participant generalisation remained challenging, with all pose-based models achieving only 40\%–43\% accuracy—barely above the 33\% chance level. In this setting, task performance metrics generalised somewhat better (52\%), and combining them with pose features yielded modest improvements (51\%–53\%). However, even these combined models remained well below the within-participant ceiling. This pattern aligns with prior work showing that workload indicators are highly idiosyncratic, often requiring calibration \citep{baldwin_adaptive_2012,walter_using_2013,cinaz_monitoring_2013}. The poor cross-participant transfer likely reflects substantial inter-individual differences in how people move, position themselves, and engage with tasks—variability that may be amplified in our relatively homogeneous sample of university students. A larger and more diverse dataset might reveal more generalisable patterns or enable transfer learning approaches that adapt to new individuals with minimal calibration data.

The participant-specific learning curves revealed both the promise and challenge of individualised workload monitoring. Pose kinematics improved continuously from 49.5\% accuracy with minimal calibration (2 minutes per condition) to 73\% with extended calibration (11 minutes per condition)—a 23.5 percentage point gain with no plateau—suggesting these features capture rich hierarchical structure in movement dynamics. Task performance metrics, conversely, improved only 6.1 percentage points and plateaued around 50\%. Individual variability was substantial: at maximum training size, accuracy ranged from 23\% to 98\% (SD $>$ 20\%), likely reflecting genuine differences in how individuals express workload. Critically, baseline calibration data provided no predictive value (37\% accuracy) and degraded performance when combined with in-task data, underscoring that workload expression is context-dependent rather than a stable trait. This necessitates task-embedded calibration for reliable monitoring.

Future work should address cross-participant generalisation through transfer learning \citep{pan2009survey,weiss2016survey} and meta-learning approaches \citep{hospedales2021meta} that initialise models on population data before fine-tuning to individuals. Domain adaptation techniques \citep{ganin2016domain,wang2018deep} could disentangle shared load-related dynamics from idiosyncratic movement styles, while deep learning architectures operating on raw video \citep{simonyan2014two} may discover richer representations than hand-crafted features. Temporal models such as Transformers \citep{vaswani2017attention,cordonnier2021differentiable} could better capture long-range dependencies, and multi-modal fusion \citep{baltruvsaitis2018multimodal} integrating pose with keyboard dynamics, mouse movements, or pupillometry may improve robustness. Expanding to diverse populations, professional domains, and naturalistic settings would reveal whether these patterns generalise beyond controlled laboratory multitasking.

Practically, these results support real-time, off-body workload monitoring using commodity cameras. Compared to eye-trackers or wearables, video-based pose tracking is unobtrusive and scalable, making it well-suited to naturalistic settings with minimal setup. While cross-participant models showed limited generalisation, the strong within-participant performance (85\%) demonstrates that with brief per-user calibration, pose-derived signals can enable adaptive interfaces that detect rising load and adjust support in real time—capturing not just whether a user is performing, but how they coordinate, move, and adapt under changing demands. 

\section{Methods}\label{sec11}

\subsection{Participants}
Seventy-two undergraduate students (52 women, 20 men; \textit{M}=21.0 years, \textit{SD}=3.51, range = 18–39) were recruited from Macquarie University's Psychology participant pool. Participants were randomly assigned to one of six counterbalanced task orders (12 per order). All gave informed consent and received course credit. The study was approved by Macquarie University's Human Research Ethics Committee.

\subsection{Task Design}
Participants completed a modified version of OpenMATB—an open-source multitasking simulation based on NASA’s Multi-Attribute Task Battery (MATB) \citep{cegarra_openmatb_2020,comstock_jr_multi-attribute_1992}. MATB simulates aviation-like workload and is widely used in cognitive load research \citep{pontiggia_matb_2024,gugerell_studying_2024,chai_examining_2024}. Version 1.3.0 was used, comprising four simultaneous tasks: system monitoring, tracking, communications, and resource management.

\subsubsection{OpenMATB Components}
\paragraph{System Monitoring} Participants monitored two indicator lights and four dynamic gauges for deviations from a predefined “normal” state (Figure~\ref{fig:matb-method}). Normally, the left light was on (green) and the right light off (grey). If the left light turned off or the right turned on (red), participants pressed F5 or F6, respectively, to restore the default state. Below, four gauges fluctuated around a central setpoint; deviations toward the edges required correction using keys F1–F4. Timely corrections (within 10 seconds) triggered a green border, while delayed or missed responses triggered a red one.

\paragraph{Tracking} Using a joystick, participants attempted to keep a moving cursor centered within a white target area (Figure~\ref{fig:matb-method}.A). Cursor colour changed from black to red if it left the target zone, providing immediate feedback. Cursor motion followed a fixed trajectory generated by the sum of sinusoidal forcing functions along the x- and y-axes, ensuring consistent difficulty across participants.

\paragraph{Communications} To align more closely with real-world aviation tasks, the standard OpenMATB communications task was simplified to focus on verbal repetition. Participants heard verbal prompts indicating a call-sign and a target frequency (e.g., “COM1, 120.5 MHz”). If the call-sign matched their assigned one, they repeated the frequency aloud by pressing and holding either a joystick button or the space bar, which initiated an audio recording (Figure~\ref{fig:matb-method}.A). Recording continued as long as the button was held, after which the system automatically saved the file and provided generic feedback (“Roger that”), regardless of accuracy.

\paragraph{Resource Management} Participants managed fuel levels in two primary tanks (A and B) by toggling pumps to transfer fuel from auxiliary tanks (Figure~\ref{fig:matb-method}.A). Fuel leaked at a constant rate, and participants used number keys to switch pumps on or off to maintain tank levels within a target range. Arrows indicated fuel flow direction, and a real-time display showed current levels and pump status, requiring ongoing adjustment to avoid overflow or depletion.

\subsubsection{Cognitive Load Manipulations}
Cognitive load was manipulated within subjects across three levels: low, moderate, and high. Each participant completed all three conditions during both a baseline phase (three 2-minute blocks) and an experimental phase (three 8-minute blocks). The order of conditions was fully counterbalanced and held constant across phases. Participants were not informed of load levels, and all scenario events were dynamically generated to avoid repetition across individuals.

Load variation was achieved by adjusting parameters across the four concurrent tasks. In the system monitoring task, higher load levels produced more frequent anomalies in gauges and indicator lights, increasing the monitoring burden. In the tracking task, target radius size decreased with load, requiring more precise joystick control. For the communications task, participants responded verbally to frequency prompts only when their callsign was mentioned. Although the prompt rate increased with load, the proportion of target to distractor prompts remained constant at 50\%. In the resource management task, the rate of fuel leakage was increased at higher loads, demanding more frequent and timely pump adjustments to prevent overflow or depletion.

These parameters—anomaly frequency, tracking precision, communication rate, and fuel consumption—were automatically adjusted via scenario files generated by custom Python scripts, ensuring consistent and reproducible manipulations across participants.

\subsubsection{Procedure}
Before the baseline trials, participants watched a short instructional video outlining the four subtasks. To confirm comprehension, they then verbally explained each task back to the experimenter. This was followed by a practice phase: each subtask was completed individually for one minute, then all tasks were performed together for two minutes at the low load level. Performance data were recorded throughout, and participants needed to reach 80\% accuracy before advancing. A short break (up to five minutes) was offered at the end of the practice.

The baseline assessment comprised three two-minute blocks—one per load level—presented in a counterbalanced order. Each block was separated by a brief setup interval. LabRecorder captured continuous video and task event data during this phase.

The experimental assessment followed the same counterbalanced structure, with three eight-minute blocks corresponding to the low, moderate, and high load conditions. After each block, participants completed the NASA-TLX workload questionnaire. 

\subsection{Materials}
Participants completed the study seated in an individual testing cubicle equipped with a Thrustmaster T.16000M joystick, HP X27q monitor, keyboard, mouse, and noise-cancelling headphones (Figure~\ref{fig:matb-method}).

Facial and upper-body movement were recorded using a Logitech Brio HD webcam (1920×1080, 60 Hz), controlled via Open Broadcaster Software (OBS v30.1.2). OBS simultaneously captured the participant’s webcam feed and screen display, storing them side by side for analysis.

Task performance data were logged by the OpenMATB software in trial-level CSV files, including accuracy and response time. Both data streams—OpenMATB and OBS—were synchronised using LabRecorder 1.16.2, producing a unified XDF file for downstream analysis.

\subsection{Task Performance}
Performance logs were exported from OpenMATB as timestamped CSV files containing event markers, task outputs, keypresses, and joystick inputs. Events for the four subtasks—tracking, resource management, system monitoring, and communications—were parsed using a custom Python script. All events were time-aligned to scenario onset.

For each subtask, a point-accuracy metric was computed by dividing correct responses (or time in target) by the total number of relevant events. Tracking accuracy was defined as the proportion of samples where the cursor was inside the target region. Resource management accuracy reflected the proportion of time both dials (A and B) were within tolerance. System monitoring accuracy was computed as hits minus false alarms, normalised by total signal detection events. Communication accuracy was defined as the proportion of appropriate keypress responses (within 15 seconds) to “own” radio prompts, minus false alarms to “other” prompts.

All metrics were calculated within overlapping windows (60-second bins with 50\% overlap).

\subsection{Pose Dynamics}

Video recordings were sampled at 60 Hz and analysed offline with OpenPose’s 70-keypoint face model \citep{cao_realtime_2017}. For each frame, OpenPose returned the \(x\)–\(y\) coordinates and a confidence score for each landmark.

\subsubsection{Preprocessing Pipeline} 

A eight-stage preprocessing pipeline was applied to ensure data quality. First, raw OpenPose data were loaded and filtered to retain only anatomically relevant keypoints for subsequent analysis. Second, landmarks with confidence scores below $0.30$ were flagged as unreliable and masked. Third, short gaps of sixty or fewer consecutive frames (1 second) in the resulting time series were filled using linear interpolation. Fourth, a zero-phase 4th-order Butterworth low-pass filter with a 10 Hz cutoff was applied to contiguous data segments to reduce high-frequency noise while preserving movement dynamics. Fifth, coordinates were normalised to screen dimensions (2560x1440 pixels) by dividing x-coordinates by screen width and y-coordinates by screen height, yielding values in the range [0,1].

\subsubsection{Template Construction}

Sixth, a global reference template was constructed using four stable facial landmarks: the nose tip (keypoint 30), side of nose (keypoint 31), top right eyelid (keypoint 37), and bottom left eyelid (keypoint 46). These landmarks provide a minimal set of reference points distributed across the face to capture head pose. The template was created by averaging positions of these four keypoints across all frames from all participants, creating a single neutral reference configuration. In addition to the global template, per-participant templates were constructed using the same four landmarks but averaging only within each participant's data.

\subsubsection{Coordinate Stabilisation and Feature Extraction} 

Seventh, three approaches to head movement stabilisation were evaluated: Procrustes alignment to the global template, Procrustes alignment to per-participant templates, and no stabilisation using screen-normalised coordinates directly. For the Procrustes approaches, each frame was aligned to its respective template using Procrustes superimposition \citep{gower1975generalized,rohlf1990extensions}, which finds the optimal rigid transformation—consisting of translation, rotation, and anisotropic scaling—that minimises the squared distance between corresponding landmarks. This alignment removes gross head translation, rotation, and size variations while preserving within-face deformations. The transformation parameters themselves (translation components, rotation angle, and scale factors) were extracted as features representing overall head movement, while the aligned landmark coordinates were used to compute features reflecting facial expression dynamics.

From the transformation parameters, head movement features were computed including the Euclidean magnitude of translation ($\sqrt{tx^2+ty^2}$), the rotation angle extracted from the rotation matrix, and the separate x and y scale factors. An additional combined head motion magnitude was calculated incorporating both translational and scaling components.

From the aligned landmark coordinates (or screen-normalised coordinates for the no-stabilisation approach), facial expression features were extracted using anatomically coherent landmark groups. 

Blink aperture was computed as the vertical distance between upper and lower eyelid landmarks, calculated separately for left eye (landmarks 38–39 versus 41–42) and right eye (landmarks 44–45 versus 47–48) and then averaged. Mouth aperture was measured as the Euclidean distance between upper and lower lip landmarks (63 and 67). Pupil displacement was calculated by first determining eye centres from the mean position of eye contour landmarks (37–42 for left eye, 43–48 for right eye), then computing the offset of pupil landmarks (69 for left, 70 for right) from their respective eye centres. Pupil offsets were decomposed into separate x and y components and averaged across both eyes, with an additional scalar magnitude metric computed from the Euclidean distances.

All three stabilisation approaches (global Procrustes, per-participant Procrustes, and no stabilisation) were processed in parallel, each yielding separate feature sets. Features were extracted within 60-second windows with 50\% overlap, and windows containing any missing values were excluded from subsequent analysis. All three methods yielded similar trends in pose dynamics; results reported here use the global Procrustes approach as it provides a consistent reference frame across all participants while effectively removing head movement artefacts, allowing direct comparisons of facial expression dynamics between individuals and conditions.

\subsubsection{Kinematic Derivative and Summary Statistics}

Eighth, for each extracted feature, kinematic derivatives were computed using finite differences. Velocity was calculated as the first temporal derivative and acceleration as the second temporal derivative. This yielded three related measures for each base feature: the original value, its velocity, and its acceleration. 

For each measure, nine summary statistics were calculated within each 60-second window: root mean square (RMS), mean, standard deviation, median, minimum, maximum, 25th percentile, 75th percentile, and first-order autocorrelation. While all metrics were included as inputs to the machine learning models, the main results focus on RMS and mean as these captured the most interpretable aspects of movement magnitude and central tendency.

\subsection{Statistical Analysis}

To assess the effects of load condition on task performance and pose dynamics, linear mixed-effects models (LMMs) were fit using the \texttt{lme4} and \texttt{lmerTest} packages in R \citep{bates2015fitting,kuznetsova2017lmertest}. All models were fitted to windowed data (60-second windows with 50\% overlap) with condition coded as an ordered factor (Low, Moderate, High) and window index mean-centred prior to inclusion. For each dependent variable (task performance metrics and pose-derived features), the model specification included fixed effects of condition and window index, plus random intercepts and slopes for window index by participant: \texttt{DV $\sim$ condition + window\_index$_c$ + (1 + window\_index$_c$ || participant)}. The double-bar notation indicates uncorrelated random effects. Random slopes for window index were included to account for participant-specific trends across time. If the model was singular or variance components approached zero, the model was automatically reduced to random intercepts only: \texttt{(1|participant)}. Degrees of freedom and $p$-values were estimated using Satterthwaite approximations. 

Pairwise comparisons between load conditions were computed using estimated marginal means (\texttt{emmeans} package) with Tukey adjustment for multiple comparisons. Effect sizes are reported as standardised mean differences (Cohen's $d$), calculated by dividing raw pairwise contrasts by the model residual standard deviation. For each contrast, the effect size represents the difference between higher and lower load conditions (e.g., $d_{M-L}$ indicates Moderate minus Low)

Models were fitted separately for each feature within each analysis domain (task performance, linear pose kinematics, and RQA metrics).

\subsection{Recurrence Quantification Analysis}
To examine nonlinear dynamics in facial movement, Recurrence Quantification Analysis (RQA) was applied to pose-derived time series. Prior to analysis, each coordinate stream was linearly rescaled to the unit interval [0, 1] to prevent any marker or axis from disproportionately influencing the Euclidean distance used in state-space reconstruction.

The time delay $\tau$ was determined using Average Mutual Information (AMI). Rather than selecting the first local minimum, the first plateau was chosen—a more robust criterion for noisy, quasi-periodic data such as short-term human movement \citep{fraser_independent_1986, wallot_calculation_2018, nishimoto_implicit_2024}. To estimate the embedding dimension $m$, a False Nearest Neighbours (FNN) analysis was conducted. The proportion of false neighbours dropped below 1\% at $m = 4$, consistent with prior work on postural and gait dynamics at similar sampling rates \citep[e.g.,][]{bruijn_statistical_2009}. This value was adopted for all auto- and cross-recurrence reconstructions to maintain comparability. For all time series, $\tau = 20$ and $m = 4$ were used. Representative AMI and FNN plots are provided in Supplementary Materials.

The recurrence threshold $\epsilon$ was set as 20\% of the mean pairwise distance ($\epsilon = 0.20\overline{d}$), for auto-recurrence analysis and 30\% ($\epsilon = 0.30\overline{d}$) for the cross-recurrence analysis between gaze and head movements, yielding recurrence rates in the recommended 2–5\% range. This fixed-percentile approach avoids over-sensitivity to $m$ \citep{kraemer_recurrence_2018, nkomidio_recurrence-based_2022}. While some features (blink and mouth) had slightly lower or higher recurrence rates at 20\% distance, they were kept for comparison. Different radii were tested and showed the same trends. To suppress trivial autocorrelations, a Theiler window of 2 was applied. The minimum diagonal line length was set to $l_{\text{min}} = 4$, filtering out brief, noisy recurrences while preserving structure. Although lower values (e.g., $l_{\text{min}} = 2$) are sometimes used for physical systems, pilot tests showed they artificially inflated determinism, making the measure less sensitive to load-related changes \citep{babaei_selection_2014}.

RQA was computed in sliding windows of 60 seconds with 50\% overlap (3600 samples per window at 60 Hz). This length meets the $\geq$1000-sample criterion for stable recurrence distributions \citep{marwan_recurrence_2007}, and aligns with recommendations for human movement analysis \citep{wallot_multidimensional_2016}.

Both auto RQA and cross RQA were carried out using a Nonlinear Time Series Python toolbox \citep{macpherson2024advanced}. The primary recurrence-based dependent variables (DVs) extracted from each matrix are defined in Table~\ref{tab:rqa_dvs} and include metrics such as recurrence rate, determinism, entropy, complexity, and divergence, among others.

\begin{table*}[h!]
\centering
\caption{Recurrence-based dependent variables (DVs) computed.}
\label{tab:rqa_dvs}
\begin{tabular}{ll}
\toprule
\textbf{DV} & \textbf{Definition / Computation} \\
\midrule
\textbf{Recurrence Rate (RR)} & \% of points in the recurrence matrix with distance $\leq \epsilon$ \\
\textbf{Determinism (DET)} & \% of recurrent points forming diagonal lines $\geq l_{\text{min}}$ \\
\textbf{Mean Line Length} & Mean length of diagonal lines $\geq l_{\text{min}}$ \\
\textbf{SD Line Length} & Standard deviation of diagonal line lengths $\geq l_{\text{min}}$ \\
\textbf{Entropy} & Shannon entropy of the diagonal line length distribution \\
\textbf{Complexity} & Residual information: max entropy minus observed entropy \\
\textbf{Divergence} & Inverse of the length of the longest diagonal line ($1/L_{\max}$) \\
\textbf{Trend} & Linear slope of recurrence density across lower and upper diagonals \\
\textbf{Laminarity (LAM)} & \% of recurrent points forming vertical lines $\geq l_{\text{min}}$ \\
\textbf{Trapping Time (TT)} & Mean length of vertical lines $\geq l_{\text{min}}$ \\
\textbf{Vmax} & Length of the longest vertical line \\
\bottomrule
\end{tabular}
\end{table*}

\subsection{Supervised Machine Learning}

To assess whether behavioural and pose-derived features could predict cognitive load, we trained Random Forest classifiers to distinguish between Low, Moderate, and High workload conditions. Classification was performed on 60-second windows with 50\% overlap, using three types of features: (1) task performance metrics (subtask accuracy and reaction time), (2) linear pose kinematics (velocity, acceleration, displacement), and (3) nonlinear dynamics from recurrence quantification analysis.

Random Forest models were implemented using \texttt{scikit-learn} \citep{pedregosa_scikit-learn_2011} with 300 trees and balanced class weights to account for any imbalances across conditions. Unless otherwise noted, all reported accuracy values refer to balanced accuracy—the average of per-class accuracies—which provides a fair assessment when class sizes differ.

\subsubsection{Validation Strategies}
We evaluated models using two complementary approaches. The first approach tested within-participant generalisation by randomly splitting all windows into 80\% training and 20\% testing sets while maintaining class proportions. This allowed the model to learn from some windows of each participant and predict others. The second approach tested cross-participant generalisation using leave-one-participant-out cross-validation (LOPO-CV), where each of the 72 participants was held out once as the test set while training on all remaining participants. This stricter test reveals whether patterns learned from one group of people transfer to new individuals. The random-split approach was repeated 15 times with different random splits to ensure stable estimates, while LOPO-CV was repeated 15 times per participant to account for random variation in training.

\subsubsection{Feature Selection and Hyperparameter Tuning}
To identify the most informative features and optimise model performance, we employed a two-stage approach for the random-split validation. However, for LOPO-CV, we omitted feature selection to avoid data leakage. In LOPO-CV, selecting features using any subset of participants would introduce information about those participants into the model, violating the assumption that the test participant is completely unseen. To maintain the integrity of cross-participant generalisation testing, we provided all features to the Random Forest classifier and relied on its intrinsic feature importance weighting to handle irrelevant features.

For random-split models, we first removed uninformative features through aggressive filtering: features with near-zero variance (less than $10^{-8}$) were dropped as they provided no discriminative information, and when pairs of features were highly correlated (above 0.95), one was removed to reduce redundancy.

Second, we applied backward elimination to identify the minimal feature set that maintained high performance. Starting with all remaining features, we iteratively removed the least important 20\% based on permutation importance—a robust measure that reflects how much prediction accuracy drops when a feature's values are randomly shuffled. This process continued until removing more features degraded performance or fewer than 5 features remained. Permutation importance was averaged across 3 repetitions and 5 cross-validation folds for stability.

Hyperparameter tuning was initially explored to optimise the Random Forest configuration (tree depth, minimum samples per split, number of features per split, and forest size) using a random search over 30 parameter combinations evaluated with 5-fold cross-validation. However, preliminary analyses showed that tuning provided negligible performance gains while substantially increasing computational cost. We therefore used default Random Forest parameters (300 trees, balanced class weights, unrestricted tree depth) for all final analyses.

Feature selection was performed only once using the training data from the first random seed. This one-time optimisation prevented over-fitting to particular data splits while substantially reducing computation time. The selected features were then held constant across all 15 evaluation seeds.

\subsubsection{Model Evaluation}

For each validation fold (random seed for random-split, or participant for LOPO-CV), we trained a Random Forest model on the available features after standardising them to have zero mean and unit variance. The trained model generated predictions for the held-out test set, which were evaluated using multiple metrics: balanced accuracy (our primary metric), weighted F$_1$ score (harmonic mean of precision and recall), and Cohen's $\kappa$ (agreement beyond chance). We also computed precision, recall, and F$_1$ scores separately for each load level. Confusion matrices were calculated and normalised by the true number of samples in each class, expressing classification patterns as percentages. Detailed results for all models are reported in Supplementary Materials.

Final results for random-split validation represent the mean and standard deviation across 15 random seeds, while LOPO-CV results represent the mean and standard deviation across 15 random seeds per participant. All analyses were conducted in Python 3.9 using \texttt{scikit-learn} 1.0.2, \texttt{pandas} 1.4.2, and \texttt{numpy} 1.22.3.

\subsubsection{Participant-Specific Learning Curves}

To determine minimum calibration requirements for participant-specific workload detection, we trained individual models for each participant using varying amounts of training data. Training sizes ranged from 2 to 11 windows per condition (Low, Moderate, High), with each window representing 60 seconds of data. At each training size, we sampled the first $N$ windows from each condition in order, reserved the subsequent window as a buffer (to avoid testing on data with 50\% overlap with training data), and used all remaining windows for testing.

For models including baseline calibration, training size 0 used only the three baseline windows (3 per condition, collected during the 2-minute baseline blocks) and tested on experimental windows. Training sizes 2-11 included baseline windows plus the specified number of experimental windows per condition. Models without baseline calibration used only experimental windows.

Each participant-specific model was trained using the same Random Forest configuration as the main analyses (300 trees, balanced class weights), with features standardised within each participant. Feature selection via backward elimination was performed once per participant at the largest training size (11 windows per condition) and applied consistently across all smaller training sizes for that participant. To ensure robust estimates, we repeated the training-test split procedure 10 times with different random seeds for each participant and training size combination, then averaged performance metrics across seeds. 

Learning curves were generated for three feature sets: (1) task performance metrics only, (2) linear pose kinematics only, and (3) combined pose and performance features. Additional analyses tested the contribution of baseline calibration and recurrence quantification analysis features. Results represent the mean and standard deviation of balanced accuracy across participants at each training size, with detailed metrics reported in Section 4 of the Supplementary Materials. 

\textbf{Ethical approval declarations} All experimental protocols were approved by the Macquarie University Human Research Ethics Committee. Informed consent was obtained from all participants prior to participation.

\backmatter

\bmhead{Supplementary information}

Supplementary materials include full statistical results, machine learning model outputs, and all analysis scripts, available at \\
\url{https://github.com/MInD-Laboratory/Measuring_Workload_Dynamics_in_OpenMATB}. The complete raw dataset can be accessed at \url{https://osf.io/dzgsv/}.

\bmhead{Acknowledgements}

We would like to thank Zac Stritch-Hoddle, Edlyn Kim, and Patrick Nalepka for their contributions to related research. This research was supported by funding from the Defence Science and Technology Group (DSTG), Australian Department of Defence, via the Human Performance Research Network (HPRNet) and the Centre for Advanced Defence Research in Robotics and Autonomous Systems (CADR-RAS).

\bmhead{Author contribution}
The study was conceived and designed by all authors. Data were collected by C.S., M.J.G., and J.W. Data analysis was undertaken by C.S. and M.J.R. Preparation of the manuscript was carried out collaboratively by all authors. All authors reviewed, provided critical feedback on, and approved the final version of the manuscript.

\newpage
\twocolumn
\bibliography{references_cleaned}

\end{document}